# Energy-Efficient Infrastructure Sensor Network for Ad Hoc Cognitive Radio Network

Muhammad Usman, Dongsoo Har, *Senior Member, IEEE,* and Insoo Koo, *Member, IEEE*

*Abstract*—We propose an energy-efficient network architecture that consists of ad hoc (mobile) cognitive radios (CRs) and infrastructure wireless sensor nodes. The sensor nodes within communications range of each CR are grouped into a cluster and the clusters of CRs are regularly updated according to the random mobility of the CRs. We reduce the energy consumption and the end-to-end delay of the sensor network by dividing each cluster into disjoint subsets with overlapped sensing coverage of primary user (PU) activity. Respective subset of a CR provides target detection and false alarm probabilities. Substantial energy efficiency is achieved by activating only one subset of the cluster, while putting the rest of the subsets in the cluster into sleep mode. Additional gain in energy efficiency is obtained by two promising propositions : selecting nodes from the active subset for actual sensing and switching the active subset to sleep mode by scheduling. The sensor nodes for actual spectrum sensing are chosen considering their respective time durations for sensing. Even the only active subset is switched to sleep mode for a certain number of time slots, utilizing the history of PU activity. We compare the proposed CR network with existing approaches to demonstrate the network performance in terms of the energy consumption and the end-to-end delay.

## I. INTRODUCTION

ACCORDING to the Federal Communications Commission (FCC), utilization of the statically assigned spectrum varies from 15% to 85%, depending upon spatio-temporal variations [1, 2]. In order for a secondary user, which cannot be active when the primary user (PU) is active, to utilize the spectrum licensed to a PU, the activity of the PU should be closely monitored [3]. One possible approach is to use cognitive radio (CR) transceivers for spectrum sensing and sending their observations to a fusion center to determine the presence of the PU signal [4, 5]. However, this approach incurs high cost and high energy consumption.

A more appealing approach is to perform sensing via cost-effective and dedicated sensor network [6, 7]. Use of the sensor network for spectrum sensing is being explored by regulatory bodies like the FCC, which has invited experts to draft proposals for the use of a sensor network with low cost/energy/delay for enhanced spectrum sensing [8]. Energy-efficient spectrum sensing by a sensor network offers advantages such as more effective detection of a weak PU signal (by location diversity of the sensor nodes) and better protection of the PU due to high reliability in detection. Furthermore, this approach is more appropriate for mobile CRs where cooperative spectrum sensing is more difficult in the absence of a fusion center and cooperation between the CR users cannot be easily achieved. However, there are still certain challenges/disadvantages in such a network, which are yet to be resolved; examples are ownership of the sensor network, information dissemination by the sensor network, usage fees, etc.

The sensor network required for spectrum sensing should be a low-cost network consisting of a large number of spatially distributed sensor nodes equipped with sensing, processing, and communications capabilities. Because the sensor nodes are characterized by their limited resources (e.g., storage capacity and processing power, and typically non-replaceable, limited-capacity batteries), efficient consumption of the energy, which affects network lifetime, is a major concern. The sensor nodes carry out spectrum sensing by means of energy detection, and report the results to the CR acting as a fusion center. The decision fusion at the CR employs the OR-rule, which decides the presence of the PU signal when at least one of the sensor nodes reports its presence. The effect of location diversity is more profound with more sensor nodes involved in spectrum sensing.

In this paper, a CR network (CRN) with disjoint subsets for each cluster of sensor nodes is proposed as a solution to the problem — effective sensing achieved with high energy efficiency. The CRN is composed of ad hoc CRs, assigning mobility to CRs to be more general, and infrastructure sensor nodes. An ad hoc CR, which is a cluster head, is surrounded by a cluster of infrastructure sensor nodes within one-hop communication range of the CR, and each cluster is further partitioned into subsets. To achieve energy efficiency, sleep–wake scheduling for the subsets based on the statistical behavior of the PU is also proposed. Relevant procedures for effective sensing achieved with high energy efficiency are as follows.

Step 1: Divide the whole sensor network into clusters (cluster formation) and update the clusters if any CR moves (cluster updating).

Step 2: Divide each cluster into disjoint subsets (subset formation).

Step 3: Minimize sensing energy by taking variable sensing time into consideration, because a sensor node receiving a signal with high signal-to-noise ratio (SNR) consumes less energy and takes less time in sensing, which means it goes to sleep for a longer time to save energy.

Step 4: Schedule the subsets by activating only one of the subsets while switching the others to sleep mode. The PU's historical activity data are used to determine the sleep time of the subsets. These procedures lead to energy-efficient sensor network operation for the CRN, which is shown in Section VI by comparisons with the SENDORA network and a CRN with the LEACH-C protocol.

Main contributions of this paper are as follows.

1. We proposed an energy-efficient cluster updating and subset formation (CUSF) process for the operation of ad hoc CRs assisted by an infrastructure sensor network. The CRs randomly move in time and the subsets of the clusters in the sensor network are updated accordingly. In this paper, theoretical analysis of the subset formation is also presented.

2. Only one subset in a cluster is active at a time, while the others switch to sleep mode. For further reduction of energy consumption, the actual sensor nodes for spectrum sensing are selected from the given active subset according to a separately proposed algorithm. Energy savings during spectrum sensing is a critical matter with a CRN including many sensor nodes. Most of the published works consider only communication energy or processing energy when evaluating the energy consumption of the network, so energy consumed during the sensing stage is often ignored. Though energy for each sensing is considerably less than communication energy, the short interval in the periodic sensing process of the CRNs makes it significantly important. Thus, minimization of sensing energy helps to prolong the lifetime of the sensor network.

3. Even the one active subset can be switched to sleep mode for a certain number of time slots by the proposed scheduling algorithm, based on the history of PU activity. The proposed scheduling achieves additional energy efficiency at the cost of a slightly increased error in PU detection.

4. We investigated comprehensive energy consumption of the sensor network with the proposed architecture. Overall energy consumption of the sensor network involves energy consumed in setup, sensing, sending, and sleep stages. In the literature, the energy consumed in sensor network setup has mostly been ignored. However, due to free and frequent moves of the CRs and the subsequent CUSF process for each move, energy consumed in the setup stage is additionally considered in this paper.

The remainder of this paper is structured as follows. Related work is given in Section II. System description, including subset formation, is presented in Section III. In Section IV, energy consumption during the setup stage is discussed. In Section V, energy consumption during the sensing and the sending (reporting) stages is described. Minimizing sensing energy with a minimum number of sensor nodes for actual sensing and maximizing sleep time by making use of PU's historical behavior are also discussed in this section. Simulation results in comparison with other approaches (the SENDORA network and a CRN with the LEACH-C protocol) are presented in Section VI. This paper is concluded in Section VII.

## II. RELATED WORK

Liu et al. [9] and Weiss et al. [10] proposed application of a sensor network for CR. However, they did not describe the architecture and the topology of the sensor network. Mercier et al. [11] proposed sensor-assisted CR, namely a sensor network for dynamic and cognitive radio access (SENDORA), where information on PU activity detected by a separate sensor network is transmitted via a single sink to the CRN in multi-hops. The SENDORA network addressed a comprehensive perspective of a sensor-assisted CRN. Nevertheless, it is subject to failure when the sink node breaks down, and it suffers from high energy consumption as well as high end-to-end delay because of multi-hop transmissions to the CR. Akan et al. [12] and Joshi et al. [13], described a CR sensor network (CRSN) where the conventional wireless sensor nodes are equipped with CR functionality. The CRSN requires highly complicated sensor nodes, so the high cost of a CRSN makes it impractical.

A sensor network of clusters with a hierarchical routing protocol to increase network lifetime was reported by Huang et al. [14]. They showed, with many sensor nodes, reduction of energy consumption by hierarchical routing instead of flat routing. However, their work is not related to mobile (ad hoc) CRs. Heinzelman et al. [15] proposed an energy-efficient routing protocol with low end-to-end delay, e.g., low energy adaptive clustering hierarchy (LEACH). The LEACH protocol did not, however, consider the energy state of cluster heads and sensor nodes. To enhance energy efficiency of a sensor network, various approaches have appeared. In case of wide band sensors such as orthogonal frequency division multiplexing (OFDM) type sensors, peak power reduction [16-17] is critical. Optimal sleep-wake scheduling to extend the network lifetime is investigated in [18]. However, these schemes result in increment of the packet delay as each sensor node waits for its next hop relay to wake up. Kim et al. [19] proposed packet forwarding by each sensor node to the first awake neighbor node. This method is prone to worsening the packet delay even more if the first awake node is in a direction opposite to the sink or destination node. Deng et al. [18] devised sensor scheduling by grouping the sensors into non-disjoint subsets. Each subset is activated successively to extend the network lifetime. However, subset formation does not take into account the residual energy of the nodes. Anastasi et al. [20] proposed a protocol to extend the lifetime of a sensor network by dynamically adjusting the duty-cycle of the sensor nodes. The hybrid energy-efficient distributed clustering protocol by Younis and Fahmy [21] requires neighbor nodes to exchange information, which results in increased communication overhead. Efficient scheduling of the sensor nodes was discussed in [22-24]. Vaidehi et al. [24] formed the subsets by random selection of the initial sensor node without considering its energy state. Furthermore, the number of subsets to be formed is assumed to be known a priori.

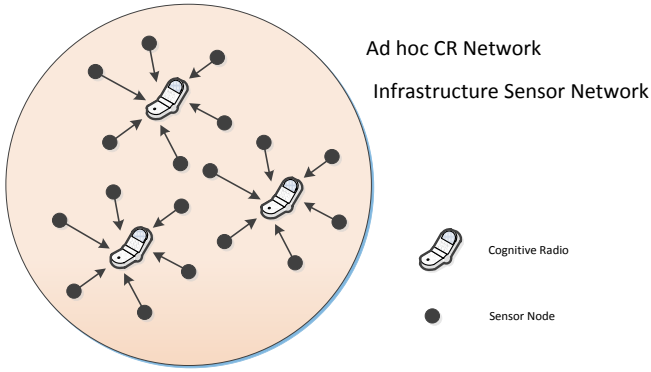

Fig. 1. Sensor-assisted CRN where CRs are surrounded by sensor nodes.

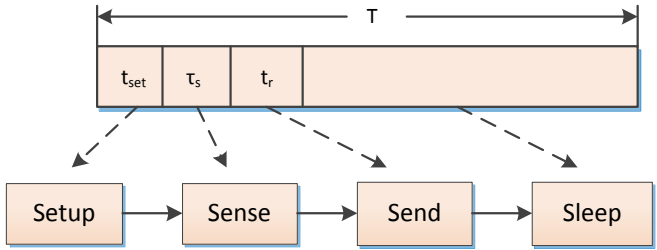

Fig. 2. Structure of a time slot for the proposed CUSF process

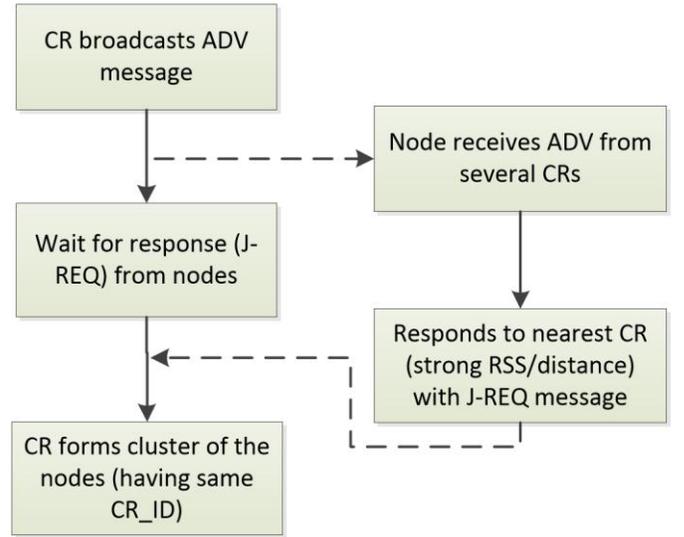

Fig. 3. Flow chart for cluster formation.

## III. SYSTEM DESCRIPTION

Consider the sensor-assisted CRN with ad hoc CRs in Fig.1. Each mobile station acts as a CR that is surrounded by sensor nodes. It is assumed that the PU operates in a time-slotted fashion. A sensor node in active mode goes through quadruple $S$-stages (setup, sense, send, and sleep) in a time slot, as shown in Fig. 2. The sensor nodes send (report) the sensing results directly to the CR that serves as a cluster head. Since an infrastructure sensor network is considered, positions of the sensor nodes are assumed to be known to the CR. It is also assumed that each sensor node knows its own position. Each CR has the ability to find its geolocation from an embedded GPS module. A CR may move in any direction within a predefined area. An error-free common control channel is assumed for the exchange of control information between nodes and the CR [25]. All the CRs have the same communication range, denoted by $r_{CR}$, and the sensor nodes have a communication range, denoted by $r_S$. It is assumed that $r_S < r_{CR}$.

### A. Cluster Formation

A CR broadcasts an advertisement (ADV) message which contains the identification number (ID) of the CR, its position, the nodes registered to the CR (Nodes), and a header field. The purpose of the header field is to differentiate the ADV message from other types of message or data. The format of the ADV message is given as follows

| Header | ID | Position | Nodes |
|--------|----|----------|-------|

Nodes within $r_S$ from the CR respond by sending a join request (J_REQ), which consists of the identification number of the node (N_ID), the identification number of the destination CR (CR_ID), the energy state (E_rem), e.g., amount of remaining energy of the node, and the SNR of the node. The format of the J_REQ is

| N_ID | CR_ID | E_rem | SNR |
|------|-------|-------|-----|

A node may receive multiple ADV messages from different CRs. In this case, the node will join the CR that is closest to it in order to consume the minimum transmission energy. Note that a node knows the position(s) of the CR(s) via the ADV messages. If a node is equi-distant from two or more CRs, it will join the CR with the smallest number of registered nodes to minimize waiting time for sending the sensing result. On receiving the J_REQ from the sensor node(s), the CR adds the node(s) to the list of registered nodes, e.g., the Nodes field in the ADV message. The flow chart of the cluster formation is given in Fig. 3. An example of cluster formation of sensor nodes with their CRs is shown in Fig. 4. C$i$, $i$=1,...,4, indicates the number of sensor nodes registered to the $i$-th CR. The nodes are grouped into clusters 1, 2, 3, and 4, respectively. The empty circles in the figure represent unclustered nodes because of their locations outside the communication ranges of the CRs. In the figure, the clusters are formed (updated) according to the shifted CR3 and static CR1, CR2, and CR4.

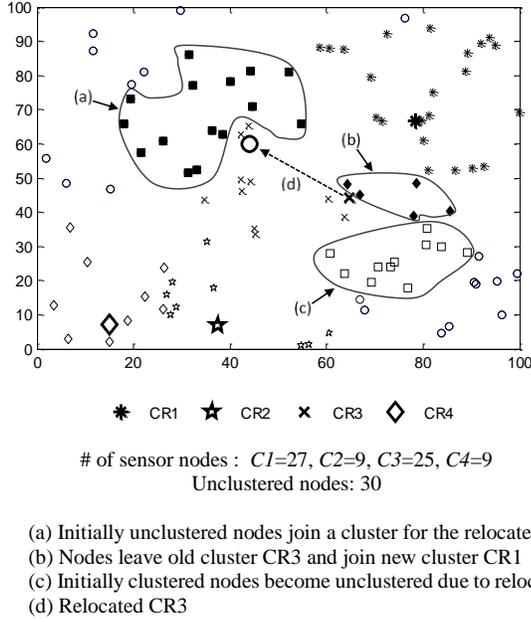

# of sensor nodes : *C1*=27, *C2*=9, *C3*=25, *C4*=9
Unclustered nodes: 30

(a) Initially unclustered nodes join a cluster for the relocated CR3
(b) Nodes leave old cluster CR3 and join new cluster CR1
(c) Initially clustered nodes become unclustered due to relocated CR3
(d) Relocated CR3

Fig. 4. Cluster formation and cluster updating. *Ci* indicates the number of sensor nodes of the *i*-th cluster.

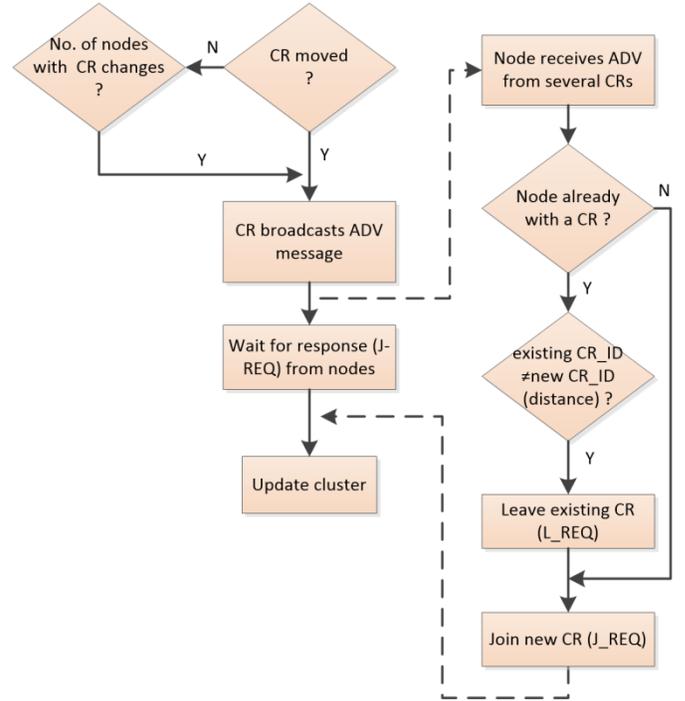

Fig. 5. Flow chart for cluster updating.

## B. Cluster Updating

It is assumed that a CR does not leave its position for time duration $t_{set} + \tau_s + t_r$ in a time slot, where $t_{set}$ is the network setup time, $\tau_s$ is the sensing time, and $t_r$ is the sending time (see Fig. 2). The cluster updating process will take place when 1) there is a change in the number of nodes registered to a CR, or 2) the position of the CR changes. When cluster updating occurs for either reason, the relocated CR initiates the update process. Unclustered nodes join the cluster of the relocated CR when they receive the ADV message from it. When a node that already joined a cluster receives the ADV message, it will leave the old (existing) cluster only if the distance to the CR of the new cluster is less than the distance to the CR of the old cluster. If the node decides to join the new cluster based on the shorter distance, the node will send a leave request (L_REQ) to the CR of the old cluster and a join request (J_REQ) to the CR of the new cluster. Format of the L_REQ message is as follows:

| N_ID | CR_ID |
|---|---|

When a CR receives an L_REQ from the registered node, the node is deregistered from the CR and the CR updates its cluster and the Nodes field in the ADV message. The cluster updating procedure is shown by the flow chart in Fig. 5. An example of cluster updating is shown in Fig. 4. CR3 moves to a new position marked by the big circle and broadcasts an ADV message. Clusters for CR2 and CR4 remain unchanged because they are outside the communication range of CR3. Due to relocation of CR3, a group of nodes initially unclustered joins the cluster for CR3 (case (a) in Fig. 4) and the other group of nodes that belonged to the cluster for CR3 join the cluster for CR1 (case (b)) and another group of nodes that belong to the old cluster for CR3 become unclustered (case (c)). As a result, *C1*=27, *C2*=9, *C3*=25, *C4*=9 after cluster updating.

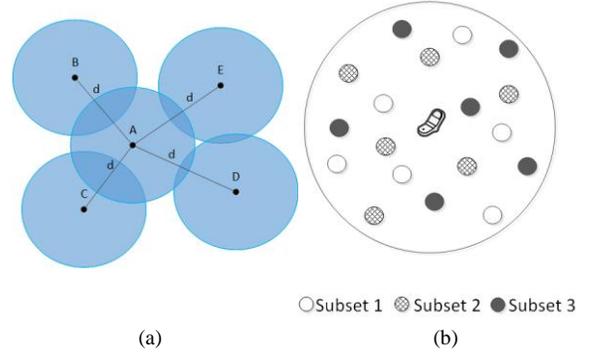

Fig. 6. Subset formation: (a) coverage overlap, (b) disjoint subsets.

## C. Subset Formation

A cluster can be decomposed into one or more disjoint subsets, as shown in Fig. 6(b). Activating only one subset of the nodes, instead of all the subsets in a cluster, significantly reduces energy consumption because deactivated subsets are switched to sleep mode. A subset of a cluster defined for this work is a group of nodes that covers the area of a cluster with minimum overlap. To avoid early failure of sensor network operation owing to the node with the least amount of energy, subset formation begins with the node that has the most remaining energy.

We need to find the number of sensor nodes in a subset that satisfies the performance criteria with minimum overlap. For example, in Fig. 6(a), coverage of four nodes overlap with the coverage of node A, and the coverage of node D is minimally overlapped with that of node A. Therefore, node D is selected by the CR as a member of the subset. For notational simplicity, the number of subsets in a cluster is denoted as $K$, and the number of sensor nodes in a cluster as $C$. To form subsets in a cluster, an algorithm similar to that of Vaidehi et al. [24] is used. While their algorithm forms subsets for a known value of $K$, the proposed subset formation algorithm does not require any prior information about $K$. Also, the number of sensor nodes, $S$, for each subset is analytically determined. The following steps are performed at the CR to form subsets of a cluster. Each subset consists of sensor nodes with minimum overlap.

*Step 1: A node with the maximum energy is selected as the starting node for subset formation.*
*Step 2: The selected node identifies all the sensor nodes (from the cluster) that have coverage overlap with that of the selected node and computes the distance between each overlapped node and the selected node to choose the one with the largest distance but which is less than $2 \times r_S$.*
*Step 3: The number of nodes in the subset is increased, whereas the number of cluster nodes is decreased by removing the chosen node from the cluster.*
*Step 4: Steps 2 and 3 are repeated with another selected node until the number of nodes in the subset becomes $S$, which is derived later.*
*Step 5: Steps 1- 4 are repeated so that K subsets are formed, and every node in the cluster is assigned to a subset. When C is not completely divisible by S, all the remaining nodes are added to the K-th subset.*

The number of nodes $S$ in a subset and the number of subsets $K$ are analytically obtained as follows.

The detection probability is defined as the probability that a sensor node correctly detects the presence of the PU signal. On the other hand, false alarm probability is defined as the probability that a sensor node incorrectly detects the presence of the PU signal when the PU signal is actually absent. The detection probability $P_{dj}$ and the false alarm probability $P_{fj}$ of the $j$-th node of a subset can be respectively given as follows [26, 27]:

$$P_{dj} = Q_u \left( \sqrt{2\gamma_j}, \sqrt{\varepsilon} \right) \tag{1a}$$

$$P_{fj} = \frac{\Gamma \left( u, \dfrac{\varepsilon}{2} \right)}{\Gamma \left( u \right)} \tag{1b}$$

where $\gamma_j$ is the SNR at the $j$-th node, $\varepsilon$ denotes the energy threshold for a local decision, $u$ represents the number of samples, $\Gamma(.,.)$ is the incomplete gamma function, $\Gamma(.)$ is the complete gamma function, and $Q_u(.,.)$ is the generalized

Marcum Q-function. Note that $\gamma_j$ is reported to the CR as a part of the J_REQ message. Since the OR fusion rule is adopted at the CR, global detection probability $Q_d$ and global false alarm probability $Q_f$ are given, respectively, as

$$Q_d = 1 - \prod_{j=1}^{S} (1 - P_{dj}) \tag{2a}$$

$$Q_f = 1 - \prod_{j=1}^{S} (1 - P_{fj}). \tag{2b}$$

$Q_d$ and $Q_f$ must satisfy the required performance level as follows:

$$Q_d \geq Q_d^{\min} \tag{3a}$$

$$Q_f \leq Q_f^{\max} \tag{3b}$$

where $Q_d^{\min}$ is the minimum global detection probability required, and $Q_f^{\max}$ is the maximum global false alarm probability allowed. Substitutions of (2a) into (3a) and (2b) into (3b) give

$$1 - \prod_{j=1}^{S} (1 - P_{dj}) \geq Q_d^{\min} \Leftrightarrow 1 - Q_d^{\min} \geq \prod_{j=1}^{S} (1 - P_{dj}) \tag{4a}$$

$$1 - \prod_{j=1}^{S} (1 - P_{fj}) \leq Q_f^{\max} \Leftrightarrow 1 - Q_f^{\max} \leq \prod_{j=1}^{S} (1 - P_{fj}). \tag{4b}$$

Let $P_d^{\min}$ be the minimum detection probability, and let $P_f^{\max}$ be the maximum false alarm probability among all the sensor nodes of the cluster. Since $P_d^{\min}$ is the minimum bound of $P_{dj}$, $j=1,...,C$ and $P_f^{\max}$ is the maximum bound of $P_{fj}$, $j=1,...,C$, $P_{dj} \geq P_d^{\min}$ and $P_{fj} \leq P_f^{\max}$, $j=1,...,C$. Hence,

$$(1 - P_d^{\min})^S \geq \prod_{j=1}^{S} (1 - P_{dj}) \tag{5a}$$

$$(1 - P_f^{\max})^S \leq \prod_{j=1}^{S} (1 - P_{fj}). \tag{5b}$$

Since equations (4a) and (4b) should be satisfied even when $P_{dj} = P_d^{\min}$ and $P_{fj} = P_f^{\max}$ for all the $j$ values, $j=1,...,C$,

$$1 - Q_d^{\min} \geq (1 - P_d^{\min})^S \tag{6a}$$

$$1 - Q_f^{\max} \leq (1 - P_f^{\max})^S \tag{6b}$$

Taking the logarithm of both sides of equations (6a) and (6b) gives

$$\left\lceil \frac{\log(1 - Q_d^{\min})}{\log(1 - P_d^{\min})} \right\rceil \leq S \leq \left\lfloor \frac{\log(1 - Q_f^{\max})}{\log(1 - P_f^{\max})} \right\rfloor \tag{7}$$

where $\lfloor \cdot \rfloor$ is the floor function and $\lceil \cdot \rceil$ is the ceiling function. $S$ is the maximum number of sensor nodes in a subset that

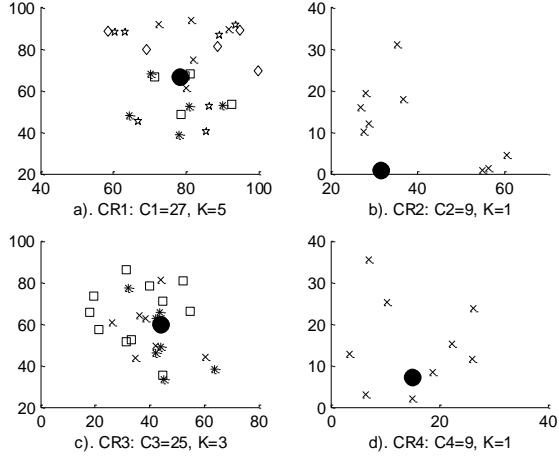

Fig. 7. Subset formation for CRs where CRs are represented by big black dots, and nodes in the subsets are denoted by different symbols. Nodes belonging to the same subset are shown with identical symbols.

satisfies the false alarm constraint [28, 29]. Hence, from equation (7), $S$ is given as

$$S = \left\lfloor \frac{\log(1 - Q_f^{\max})}{\log(1 - P_f^{\max})} \right\rfloor \qquad (8)$$

Note that $S$ is obtained from $P_d^{\min}$ and $P_f^{\max}$. Thus, any combination of $S$ sensor nodes in the cluster will satisfy the minimum global detection probability $Q_d^{min}$ and the maximum global false alarm probability $Q_f^{max}$. The number of subsets, $K$, in the cluster can be obtained by $C$ divided by $S$ as follows

$$K = \left\lfloor \frac{C}{S} \right\rfloor \qquad (9)$$

Fig. 7 shows examples of subsets in clusters. Since $S$ is derived by considering $P_d^{\min}$ and $P_f^{\max}$, the actual number of nodes ($\bar{S}$) of different subsets of a cluster for spectrum sensing varies and is always less than or equal to $S$. An algorithm to determine $\bar{S}$ from the consideration of the sensing time to achieve minimum energy consumption during the sensing stage is presented in subsection V. A.

The required level of global detection probability $Q_d^{min}$ and global false alarm probability $Q_f^{max}$ affect the performance of CRNs, e.g., interference with the PU and utilization of the spectrum, and energy consumption of sensor networks. Increasing the minimum global detection probability guarantees more protection for the PU and leads to less interference with the PU. A higher value of $Q_d^{min}$ requires a larger number of sensor nodes $\bar{S}$ for actual sensing, and as a result, increases energy consumption. On the other hand, a lower value of $Q_f^{max}$ enhances utilization of the spectrum. Lower value of $Q_f^{max}$ is preferable, but results in larger size of the subset ($S$) and also $\bar{S}$. Therefore, a lower value of $Q_f^{max}$ causes higher consumption of energy.

After subset formation, the CR selects the subset that has the maximum total energy, and creates a time division multiple access (TDMA) schedule for the nodes in the selected subset [30]. This scheduling information is transmitted to the sensor nodes in the form of beacon messages by the CR. Only the nodes included in a subset are active; all other nodes in other subsets remain in sleep mode. The order of time slots in a TDMA frame is assigned according to descending order of SNR values of the sensor nodes. Thus, the first time slot is assigned to the sensor node with the highest SNR, and so on.

## IV. ENERGY CONSUMPTION DURING SETUP STAGE

The CR nodes are assumed to be more powerful devices than sensor nodes in terms of the amount of resources. The energy consumption model by Heinzelman et al. [15] for the sensor network is considered. Free-space pathloss is considered between the nodes and the CR, due to assumed geometrical proximity of sensor nodes to the registered CR. The total energy consumed by a sensor node can be decomposed as

$$E_T = E_{set} + E_s + E_r \qquad (10)$$

where $E_{set}$, $E_s$, and $E_r$ are the energy consumed in the setup stage (setting up the cluster and the subset), the sensing stage, and the sending (reporting) stage, respectively. The processing energy at the senor nodes is ignored, because it is significantly smaller than sensing and reporting energy. In this section, energy consumed for the setup stage is discussed.

The energy consumed for the setup stage consists of the energy consumed in cluster formation, updating, and subset formation. Most of the existing protocols for various network architectures ignore clustering energy during network setup, making such protocols inadequate for implementation. During the cluster formation and updating process in the setup stage, energy is consumed in receiving the ADV messages that are broadcasted by the CR, and in transmitting the J_REQ and/or the L_REQ while responding to the relevant CRs. After receiving information about clustering from the sensor nodes, the CR performs subset formation. Energy is consumed by the sensor nodes in receiving the subset information from the CR. From these sequences, the energy consumed in the setup stage is expressed as

$$E_{set} = 2 \times E_{Rx} + E_{Tx} \qquad (11)$$

where $E_{RX}$ and $E_{TX}$ are the energy consumed in receiving and transmitting, respectively. The transmission energy is given by

$$\begin{aligned} E_{Tx} &= E_{tx-elec}(l) + E_{tx-amp}(l) \\ &= lE_{elec} + lE_{amp} \end{aligned} \qquad (12)$$

where $E_{elec}$ is the energy consumed over the unit size of the data by electronics, which depends on tasks such as digital coding, modulation, filtering, and signal spreading, and $E_{amp}$ is the amplifier energy over the unit size of the data, which depends on the distance to the relevant CR and acceptable bit error rate

---

**Algorithm 1** Minimizing the Energy Consumed in Sensing by a Subset

---

**Notations:**

$P_s$: power consumed in sensing

$S$: number of sensor nodes in a subset

$\bar{S}$: number of actual sensor nodes for sensing

$\tau_{sj}$: time duration for sensing of the $j$-th sensor node, $j=1,...,S$

$\tau_{s,max}$: maximum value of $\tau_{sj}$ for timely operation of network

$P_{dj}$: detection probability of the $j$-th node

$Q_d$: global detection probability

$Q_d^{min}$: minimum global detection probability

$\gamma_j$: signal-to-noise ratio (SNR) of the $j$-th sensor node

$E_s$: energy consumed in sensing in a subset

1. **Input:** $\gamma_j$, $j=1,...,S$
   **Output:** $E_s$

2. Sort nodes in descending order of SNR ($\gamma_j$), the first node in the sorted list has the highest SNR and the last node has the smallest SNR. $m[j]$, $j=1,...,S$ is the set of indices of the sorted entries of SNR such that $m[1]$ corresponds to the node with the highest SNR, and $m[S]$ corresponds to the node with the lowest SNR.

3. $j=1$; $E_s=0$;
   **while** $j \leq S$ AND $Q_d < Q_d^{min}$ **do**

4.     Select $m[j]$;

5.     Estimate $P_{dm[j]}$ (detection probability) for the node

6.     Calculate global detection probability $Q_d$ using OR-rule

7.     Calculate sensing energy only for the node satisfying $\tau_{sm[j]} \leq \tau_{s,max}$, and update $E_s$

8.     Increase $j$ by one

9. **end while**

10. $\bar{S} = j$;

---

at the CR, and $l$ is the size of the data. The energy consumed in receiving data is given below, ignoring the $E_{amp}$ in equation (12):

$$E_{Rx} = E_{Rx-elec}(l) = lE_{elec} \qquad (13)$$

## V. ENERGY CONSUMPTION DURING THE OPERATION PHASE

The operation phase consists of sensing, sending (reporting), and sleep stages. Since energy consumption during the sleep stage is negligible, compared to other stages, the sleep stage is not considered here. Energy efficiency in this phase can be achieved twofold: 1) minimizing energy consumption, and 2) maximizing sleep time. The first objective is achieved by an energy-efficient network minimizing the energy in sensing and reporting. On the other hand, the second objective is realized by efficient scheduling of the subsets.

### A. Minimizing energy consumption during the sensing stage

Sensing performance of the CRN is related to energy consumption of the sensor network. A higher value for the minimum global detection probability requires a larger number of sensor nodes to satisfy the performance constraint for a subset, which increases the energy consumption for sensing. Most of the published works consider only communication or processing energy when evaluating the energy consumption of the network [15, 31], so energy consumed during the sensing stage is often ignored. Even though sensing energy is considerably less than communication energy, the short interval between periodic sensing processes in CR networks makes it significantly important, so minimizing sensing energy helps to prolong network lifetime.

Time duration for sensing is a function of SNR [29]. It means, for a given detection probability and a given false alarm probability, a sensor node with a comparatively clean channel causing a higher SNR takes less time for sensing an event. The time duration for sensing by the $j$-th node in a subset can be expressed in terms of SNR, detection probability, and false alarm probability [29] as follows:

$$\tau_{sj} = \left( \frac{Q^{-1}(P_{fj}) - Q^{-1}(P_{dj})\sqrt{2\gamma_j + 1}}{\sqrt{f_s}\gamma_j} \right)^2 \qquad (14)$$

where $\tau_{sj}$, $\gamma_j$, and $f_s$ are the sensing time at the $j$-th node, the SNR at the $j$-th node, and the sampling frequency, respectively, and $Q(\cdot)$ is the complementary cumulative distribution of a standard Gaussian. Due to different SNRs at the sensor nodes, $\tau_{sj}$ varies accordingly. However, it must be less than predefined maximum value $\tau_{s,max}$ for timely operation of the network. The sensor node with a higher SNR performs sensing in less time and sleeps over a longer time to decrease energy consumption. The energy consumed by the $j$-th node in sensing is $E_{sj}=P_s\tau_{sj}$, where $P_s$ is the power consumed for sensing and $\tau_{sj}$ is the time duration for sensing. Energy consumed in sensing by a subset with $S$ sensor nodes is given as

$$E_s = \sum_{j=1}^{S} P_s \tau_{sj} \qquad (15)$$

With the constraints on energy consumption of the subset, minimizing the energy consumed in sensing by a subset is expressed as

$$\min \ E_s \qquad (16)$$

subject to

$$(i) \ Q_d \geq Q_d^{min}$$

$$(ii) \ \tau_{sj} \leq \tau_{s,max}, j = 1,...,S$$

Constraint (*i*) specifies that the cooperative detection probability should not be less than the minimum requirement $Q_d^{min}$. This constraint is on detection performance. Constraint (*ii*) requires that the time duration for sensing of a sensor node be less than or equal to sensing duration $\tau_{s,max}$ in a time slot. Note that the constraint on the global false alarm probability is satisfied as long as $\bar{S}$ is not greater than $S$.

A heuristic approach to get the actual number of sensor nodes $\bar{S}$ for sensing can be considered to solve the optimization in equation (16). The heuristic approach minimizes the sensing energy by selecting the first node with the highest SNR, following the sorting of the nodes in descending order of SNR, and then calculates the time duration for sensing to get the detection probability. The sensing process continues with the second node with the second highest SNR, and so on, until constraint (*i*) is satisfied. The pseudo code of the heuristic approach is given in Algorithm 1.

### B. Maximizing sleep time

The PU's historical behavior is considered to predict future states of the PU and to estimate the number of consecutive slots in which the subsets, including the only active subset, can be scheduled for sleep. The PU behavior is typically modeled by a two-state Markov chain, where the presence and the absence of the PU signal are modeled by busy and idle states, respectively, as shown in Fig. 8(a). According to the status of the PU, sensor nodes alternate between sleep and active states, as shown in Fig. 8(b). Considering temporal variation of PU activity, the PU tends to maintain its state once switched from the other state, i.e., the PU stays on a channel for at least a few slots after occupying it, or the PU will not occupy a channel for at least a few slots when switched to an idle state. From this viewpoint, positive correlation of the PU traffic [32–34] is considered, i.e., $p_{II} > p_{IB}$ and $p_{BB} > p_{BI}$ where $p_{ab}$ is the transition probability of the PU from state $a$ to state $b$, and subscript $I$, $B$ indicates idle state, busy state, respectively. A moving window for recording the history of PU activity is adopted. In each slot, the latest information on the PU updates the record, and the oldest information on the PU is deleted from the other side of the window.

The CR develops a history of PU activity based on the global decision in each time slot at the CR, which is based on the sensing results of the sensor nodes. However, due to errors in the sensing process of the nodes, the decision of the CR could be incorrect in some time slots. The CR overcomes the sensing errors (incorrect decisions) by receiving an acknowledgement (ACK) message from the CR receiver. The ACK message is received when the data recovered by the CR receiver are error-free. Note that the CR (transmitter) combines the individual sensing results of the sensor nodes to make a decision on data transmission to the CR receiver. During the sleep stage in Fig.2 for sensor nodes, the CR transmits its own data to the CR receiver and receives the ACK message from it. We assume that the ACK message is very short, compared to the duration of the time slot. If the ACK message is received before timeout occurs, it means that the PU is inactive and the sensing decision is correct. If the sensing decision construes the absence of the PU and the CR transmits data but does not receive the ACK message in due time, the sensing information (decision) is considered incorrect, and the history is updated with correct information.

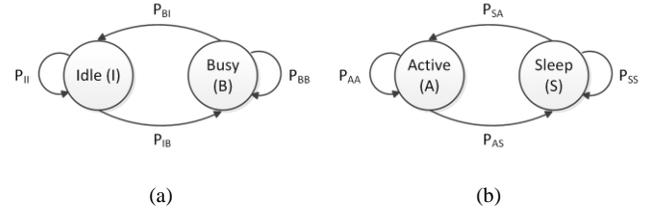

Fig. 8. Markov chain for PU and sensor node
(a) PU activity  (b) Sensor node behavior.

Consider an example of a history of PU activity within the window as follows

History of the PU: $H = 10110111011000111111101$   (17)

where '0' indicates the idle state of the PU, and '1' represents the busy state of the PU. The maximum number of consecutive (successive) slots $N_s$ in which the PU is found busy is 5 (11111). Let $G_i$, $i=2,...,N_s$, be the number of $i$-consecutive busy states. Then, the most frequent pattern of PU activity indicates $G_2=2$. If admitting that '111' is '11'+'1' and '11111' is '11'+'1'+'11,' $G_3=1$ corresponds to $G_2=1$ and $G_5=1$ matches $G_2=2$. Also, let $G_i^s$ be $G_i$ with $G_j$, $j=i+1,...,N_s$, taken into account. Then, $G_i^s \geq G_i$. For details on $G_i^s$, refer to Step 3 below. The single '1' in the streams of '1's can be matched to the 'wake' time slot of the sensor node. The algorithm to estimate the number of consecutive time slots $n_s$ for sleep is to find out the most frequent pattern of successive '1's in the PU history, and $n_s$ should be taken pessimistically, i.e., take a smaller $n_s$ if two $n_s$ values satisfy the given criterion, not to miss the chance of spectrum utilization..

The estimated number of consecutive time slots, $n_s$, in which the subsets will be switched to sleep mode, is determined as follows

*Step 1:* Determine the maximum ($N_{max}$) and minimum ($N_{min}$) number of consecutive slots in which the channel is found busy from the history of the PU.

*Step 2:* Find the number of $i$-consecutive busy slots (states) ($G_i$) that occurred. Repeat this step for $i = N_{min}$ to $N_{max}$.

*Step 3:* Find the number of occurrences of the busy state for $i$-consecutive slots in all the higher values of $i$. Occurrence of $i+1$ consecutive busy states automatically contains the non-overlapping occurrences of $i$ consecutive busy states. For example, the occurrence of three consecutive 1's contains the occurrence of two consecutive 1's. So if $i = 2$ occurred twice and $i = 3$ occurred once, then the total number of $i = 2$ that occurred is $2+1=3$. Repeat Step 3 for $i = N_{min}$ to $N_{max}$.

**TABLE 1**
Common simulation parameters.

| Description | Symbol | Value |
|---|---|---|
| Number of CRs | $M$ | 4 |
| Total number of nodes | $\sum_{i=1}^{M} Ci$ | 100 |
| Area | $A$ | $100m \times 100m$ |
| Idle probability of the PU | $P_0$ | 0.5 |
| Transition probability of PU from one state to another | $P_{IB}, P_{BI}$ | 0.3 |
| Transition probability of PU from a state to itself | $P_{II}, P_{BB}$ | 0.7 |
| Electronics energy | $E_{elec}$ | $50nJ/bit$ |
| Amplifier energy | $E_{amp}$ | $10pJ/bit/m^2$ |
| Packets | $l$ | $4000bits$ |
| Communication range of each sensor node | $r_S$ | $10m$ |
| Communication range of each CR | $r_{CR}$ | $20m$ |
| Minimum global detection probability | $Q_d^{min}$ | 0.8 |
| Maximum global false alarm probability | $Q_f^{max}$ | 0.1 |
| Time duration for sensing at node $j$ | $\tau_{sj}$ | $[0.1ms, 2.5ms]$ |
| Maximum duration for sensing | $\tau_{s,max}$ | $2ms$ |
| CR mobility range | $d_{CR}$ | $20m$ |
| Signal-to-noise ratio at node $j$ | $\gamma_j$ | $[-25dB, -5dB]$ |

$$G_i^s = G_i + \sum_{j=i+1}^{N_{max}} (G_j) \times \left\lfloor \frac{j}{i} \right\rfloor \qquad (18)$$

*Step 4: Find the probability of PU appearance, which is obtained from the number of PU appearances ($NP$) and the total number of non-zero bits ($N_1$) from the history of the PU. It contains both one-slot and multi-slot appearances of the PU.*

$$\Pr(NZ) = \frac{NP}{N_1} \qquad (19)$$

*Step 5: Find the probability of the PU appearance for i-consecutive slots, which is calculated from the number of busy states of the PU, when it stays for i consecutive slots and the total number of non-zero bits in the history of the PU.*

$$\Pr(G_i^s) = G_i^s \times \frac{i}{N_1} \qquad (20)$$

*Step 6: Compare the values obtained in Step 5 with Step 4. Select the minimum value of i that satisfies equation (21) and assign it to the number of consecutive sleeping slots $n_s$.*

$$n_s = \arg \min_{i=N_{min},\dots,N_{max}} \left[ \Pr(G_i^s) > \Pr(NZ) \right] \qquad (21)$$

From equation (21), the duration of $n_s$ slots is obtained close to the average period of the PU in busy states with various patterns in the history of PU activity. The following example will further clarify the algorithm for finding the number of consecutive sleeping slots for a subset.

**Example:**

$$H = 101101110110001111101$$

From $H$:

$NP$: Number of PU appearances = 6
$N_{max}$: Max. number of consecutive busy states (1's) = 5
$N_{min}$: Min. number of consecutive busy states (1's) = 2
$N_1$: Total busy states = 14

| No. of consecutive 1's | $G_i$ | $G_i^s$ |
|---|---|---|
| 2 | 2 | 5 |
| 3 | 1 | 2 |
| 4 | 0 | 1 |
| 5 | 1 | 1 |

$$\Pr(NZ) = \frac{NP}{N_1} = \frac{6}{14} = 0.4286$$

$$\Pr(G_2^s) = 5 * \frac{2}{14} = 0.7143$$

$$\Pr(G_3^s) = 2 * \frac{3}{14} = 0.4286$$

$$\Pr(G_4^s) = 1 * \frac{4}{14} = 0.2857$$

$$\Pr(G_5^s) = 1 * \frac{5}{14} = 0.3571$$

$$\Pr(G_2^s) > \Pr(NZ)$$

According to equation (21), the $n_s$ slots of the example is 2 slots.

Two approaches to switch the subsets to sleep mode for $n_s$ slots are considered. The first approach is to switch all the subsets to sleep mode, and the second approach is to keep the selected subset with the highest remaining energy active whereas the remaining subsets are switched to sleep mode. Since switching all the subsets to sleep mode for a certain number of slots may result in reduced utilization of the spectrum holes, it is important to ensure that during sleep mode, the opportunity for transmission is not wasted. The first approach is more energy-efficient but less reliable, while the second approach is less energy-efficient but produces more reliable sensing performance. Both approaches will be evaluated by simulation.

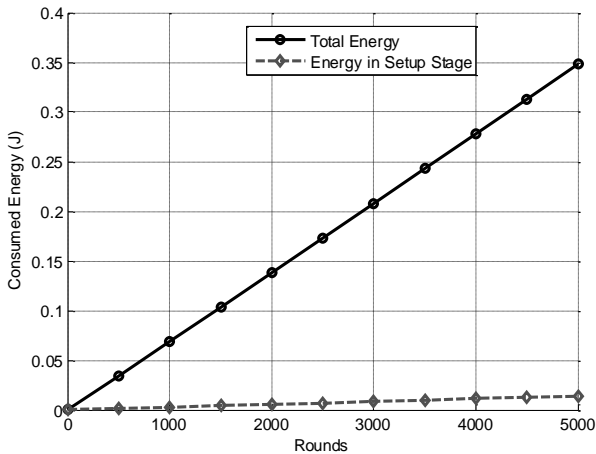

Fig. 9. Energy consumed by the proposed architecture (CUSF) in the setup stage and the setup stage plus operation phase combined.

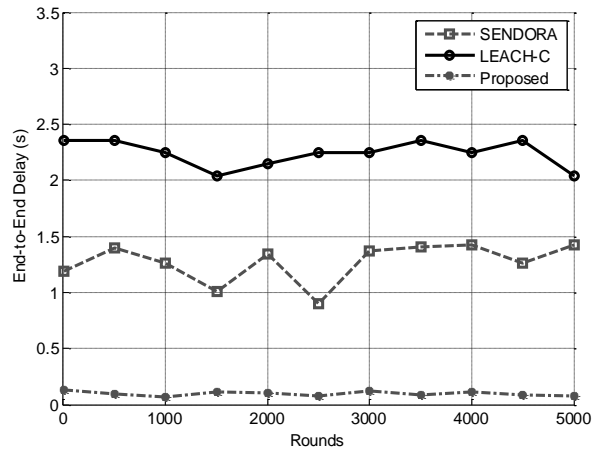

Fig. 10. Average end-to-end delay of the SENDORA network, the CRN with the LEACH-C protocol, and the CRN with the proposed architecture.

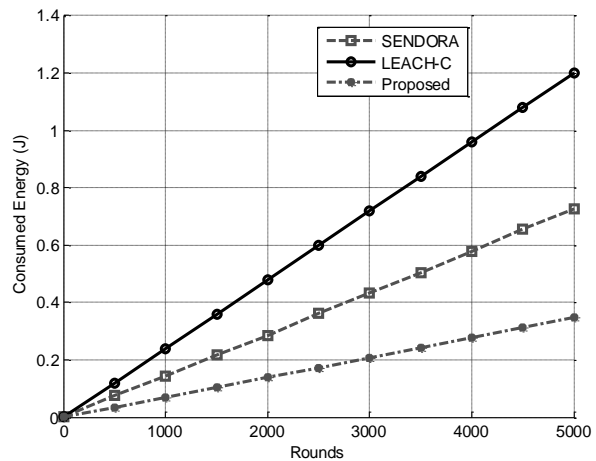

Fig. 11. Comparison of the energy consumption of the SENDORA network, the CRN with the LEACH-C protocol, and the CRN with the proposed architecture, during the setup, sensing, and sending stages.

The second and third terms in equation (10) reduce to 0 for $n_s$ slots. If there is no movement of CRs in the $n_s$ slots, the total energy consumed is reduced to 0. Subsets are switched to sleep mode for a duration of $t_{sleep}$ given by

$$t_{sleep} = n_s \times T + T_{rem} \qquad (22)$$

where $T_{rem} = T - t_{set} - \tau_s - t_r$ is the remaining time in the time slot after setup, sensing, and reporting (sending) stages.

## VI. SIMULATION RESULTS AND DISCUSSION

The common simulation parameters for the setup, sensing, and sending stages are summarized in Table 1. The CRs can move freely in any direction in the defined mobility range $d_{CR}$ within the boundaries of area $A$. The mobile nature of the CRs is modeled by the *random waypoint* (RWP) *model* [35, 36]. In the considered RWP model, a CR can move in any direction (1 to 360 degrees) by distance $d_{CR} = 20m$ in one time slot. Simulation results related to energy consumption and end-to-end delay are given in the following subsections. Lifetime of the network, described in terms of the number of rounds, is measured by the time elapsed until network energy level falls below 50%.

The typical number of subsets $K$ in a cluster varies from 2 to 4 in the simulations. $K$ tends to increase if the number of nodes in the network increases. The average number of $n_s$ for sleep is 2.2 for $P_0$=0.5. Decreasing the value of $P_0$, i.e., increasing the probability of the PU in a busy state, $n_s$ will increase. The actual number of sensor nodes $\bar{S}$ for sensing varies from 3 to 4, whereas the value of $S$ varies from 5 to 7 nodes with given conditions of simulations.

### A. Evaluation of energy consumption in the setup stage

Energy consumption of the sensor network is plotted according to the number of rounds. The number of rounds represents the number of times the CRs update their positions in the bounded region with area $A$. The consumed energy is obtained by averaging the energy consumed by all the nodes, e.g., per-node energy consumed, in the network. The initial energy ($E_0$) of each sensor node is assumed to be $5J$. Simulation is executed over 5000 rounds, and lifetime of the network in the setup stage is evaluated from the energy consumed during these rounds.

In Fig. 9, energy consumption during the setup stage is shown. The setup stage consists of the CUSF process due to the mobility of CRs, involving the exchange of control messages, e.g., ADV and J_REQ. It is evident from the figure that a negligible amount of energy ($\cong 0.3\%$) is consumed in the setup stage, compared to overall energy consumption. The lifetime of the CRN employing the CUSF process is estimated to be 85,000 rounds by the extrapolation of the plot in Fig. 9.

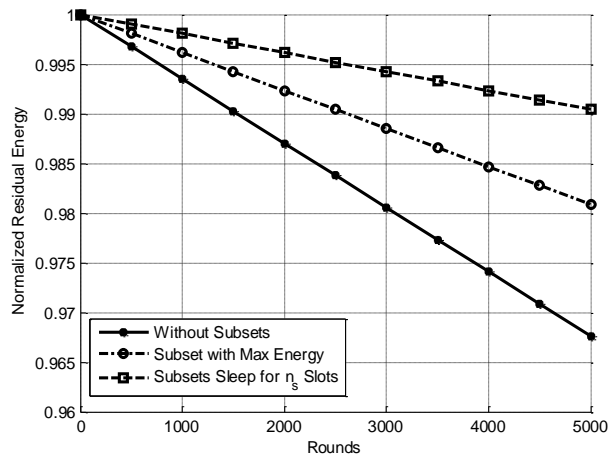

Fig. 12. Residual energy after sensing and reporting in each round with different approaches for sleep: 'subset with maximum energy', 'subsets sleep for $n_s$ slots', and 'without subsets' approaches.

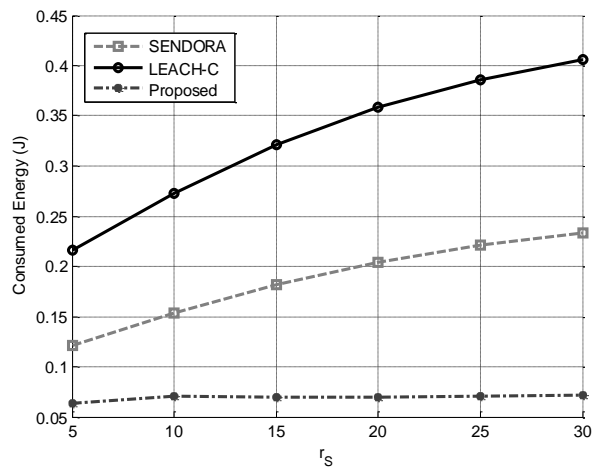

Fig. 13. Energy consumption for different communication ranges of the sensor nodes.

### B. Evaluation of end-to-end delay in the sending stage and energy consumption in the sensing/sending stages

In addition to the simulation parameters listed in Table 1, the number of iterations, sampling frequency, initial energy of each node $E_0$, and the power consumed in spectrum sensing $P_s$ are set to $5000, 300\,KHz, 5J$, and $100\,mW$, respectively.

#### 1) End-to-end delay

Figure 10 shows a comparison of the average end-to-end delay of the SENDORA sensor network, the CRN with the LEACH-C protocol, and the CRN with the proposed architecture. The end-to-end delay is defined as the time taken from sensing by the sensor nodes to the end of reporting received at the registered CR. The LEACH-C protocol uses a centralized clustering approach for the selection of cluster heads. The cluster formation (network setup), sensing, and reporting of the LEACH-C protocol are similar to those of the CUSF process. By the LEACH-C protocol, sensing is performed at the sensor nodes, and then the results are reported to the cluster head (the CR in the CRN) that aggregates the data for final forwarding to the base station. The end-to-end delay of the LEACH-C protocol comprises: i) the delay due to the formation of cluster heads by the central base station and subsequent dissemination of this information to the cluster nodes, ii) the delay at the sensor nodes caused by sensing and reporting, and iii) the delay at cluster heads from processing and transmission. The end-to-end delay of the SENDORA network is the combination of delays at the sensor nodes, the cluster heads, and the sink node. It is evident from the figure that the CRN with the CUSF process causes significantly lower delay, compared to the other protocols. The main reasons for the lower delay with the CUSF process are (i) smaller size of the subset resulting in lower transmission delay and (ii) the CR directly receiving sensing results from the sensor nodes.

#### 2) Energy consumption

Figure 11 shows a comparison of energy consumption during the setup, sensing, and sending (transmission) stages. The CRN with the CUSF process consumes the least amount of energy. That is due to having the least number of sensor nodes in sensing and sending. With the SENDORA network and the CRN adopting the LEACH-C protocol, additional energy consumption at the cluster head in aggregating data and transmitting them to the sink node for the secondary network is needed. However, the energy consumed for the additional process is comparatively negligible, so it is disregarded for comparison of energy consumption. The energy consumption of the network increases in each round because of the number of relocations of the CRs, accompanied by cluster updating and subset formation. With 5000 rounds, the CRN with the CUSF process consumes only 7% of the total energy. However, the SENDORA network and the CRN with the LEACH-C protocol consume 15% and 24% of total energy, respectively. In other words, the CRN with the CUSF process consumes 53% less energy compared to the SENDORA network and 70% less energy compared to the CRN with the LEACH-C protocol. Based on the energy consumption of the CRN with the CUSF process, the $5J$ of energy translates into 71,429 rounds, whereas it is 33,333 rounds for the SENDORA network and 24,000 rounds for the CRN with the LEACH-C protocol.

Figure 12 shows the normalized residual energy of the CRN with three different approaches for sleep. The 'subset with maximum energy' approach selects the subset with the maximum energy in a cluster for spectrum sensing in each round and keeps other subsets asleep. The approach called 'subsets sleep for $n_s$ slots' switches all the subsets in a cluster to sleep mode, including the subset with the maximum energy, for the $t_{sleep}$ duration in equation (22), which is close to $n_s$ consecutive slots, after sensing PU activity. For the 'without subsets' approach, all the sensor nodes in the cluster perform spectrum sensing. It is intuitive that the 'without subsets' approach consumes more energy and results in smaller residual energy, compared to other approaches, because all the nodes in a cluster are involved in spectrum sensing. The other approach, e.g., 'subset with maximum energy', consumes less energy than the 'without subsets' approach but more energy than the 'subsets sleep for $n_s$ slots' approach. Consequently, residual energy of the 'subset with maximum energy' approach in each round is between those of the 'without subsets' and the 'subsets sleep for $n_s$ slots' approaches.

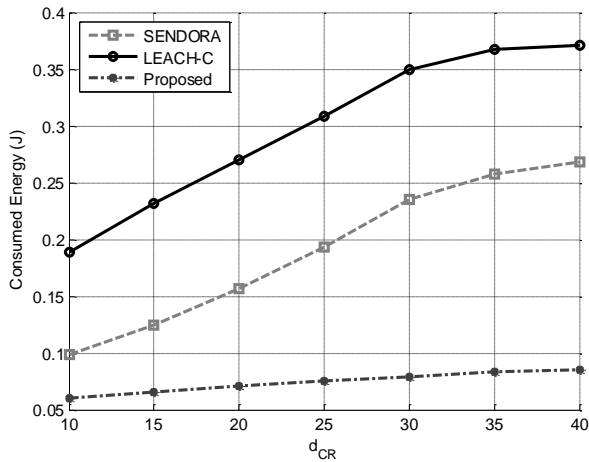

Fig. 14. Effect of the mobility distance of the CR on energy consumption of the proposed and other schemes.

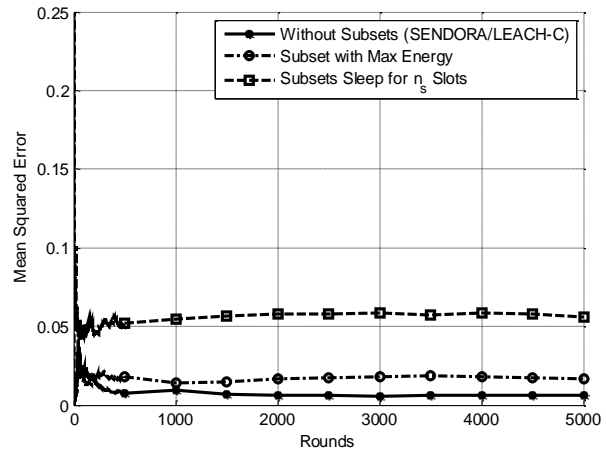

Fig. 15. Detection performance in terms of mean squared error with the 'subset with maximum energy', 'subsets sleep for $n_s$ slots', and 'without subsets' approaches. The 'without subsets' approach also refers to the SENDORA network and the CRN with the LEACH-C protocol.

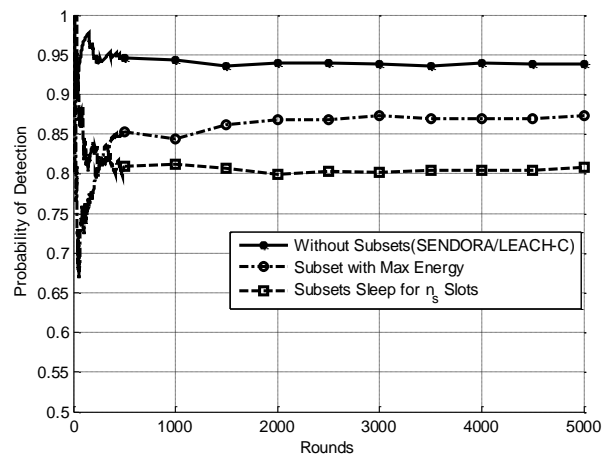

Fig. 16. Comparison of the global detection probability of the 'subset with maximum energy', 'subsets sleep for $n_s$ slots', and 'without subsets' approaches. The 'without subsets' approach also refers to the SENDORA network and the CRN with the LEACH-C protocol.

### C. Effect of various parameters on energy consumption of the network

Figure 13 shows the variation in energy consumption over various communication ranges of the sensor nodes. The increased communication range of the sensor nodes causes the unclustered nodes to join clusters of the CRs, resulting in larger cluster size, and consequently, increased energy consumption in setup, sensing, and sending in the SENDORA network and the CRN with the LEACH-C protocol. On the other hand, the larger cluster may result in a larger number of subsets of the CRN with the CUSF process, and the number of sensor nodes of the active subset is not significantly affected by cluster size. Thus, the energy consumption of the CRN with the CUSF process does not significantly depend on the cluster size.

Figure 14 shows the effect of increasing the mobility distance of the CR on energy consumption of the network. It is seen from the figure that energy consumption increases with an increased mobility distance for the CR, because moving a longer distance results in cluster formations by more sensor nodes of the relocated CR with other CRs. Similarly the relocated CR will update its own cluster by acquiring nodes from other CRs, which will further cause cluster updates in other CRs, as well. Hence, more energy is consumed in this process. On the other hand, if a CR moves a shorter distance, only few sensor nodes would be affected by its new position and thus less energy would be consumed.

### D. Evaluation of detection performance of the network with the CUSF process

Sensing performance of the sensor network is described in terms of global detection probability, global mis-detection probability, and global false alarm probability. Global detection probability is defined as the probability of detection of the PU signal by a cluster or a subset when it is actually present. Global mis-detection probability is the probability of missed detection by a cluster or a subset when the PU signal is present. Global false alarm probability is the probability of detection of the PU signal by a cluster or a subset when the signal is not present. Therefore, error in spectrum sensing comes from either missed detection or false alarm. The global mis-detection probability and the global false alarm probability can be related to the probability of error in detection as follows:

$P$(error in detection) = $P$(error in detection | PU signal is present)*$P$(PU signal is present)+ P(error in detection | PU signal is absent)*$P$(PU signal is absent)

$= (1-Q_d) * P$(PU signal is present) + $Q_f * P$(PU signal is absent) (23)

where $(1-Q_d)$ and $Q_f$ are the global mis-detection probability and the global false alarm probability, respectively.

In this paper, mean squared error (MSE) is used as a performance measure. It has statistical equivalence to the probability of error in detection. MSE is given as follows:

$$MSE_{rounds} = \frac{1}{rounds} \sum_{j=1}^{rounds} \left| H_j - D_j \right|^2$$ (24)

where $H_j$ (0 for absence or 1 for presence) represents the actual state of the PU in the $j$-th round, and $D_j$ (0 for decision of absence, or 1 for decision of presence) indicates the global decision taken by the considered approach in the $j$-th round. The 'absence' and the 'presence' correspond to the 'idle' state and the 'busy' state in Fig. 8, respectively. When $H_j$=1 and $D_j$=0, pertaining to $(1-Q_d) * P$(PU signal is present), error occurs, and when $H_j$=0 and $D_j$=1, pertaining to $Q_f * P$(PU signal is absent), the other type of error occurs. The MSE of the 'subset with maximum energy', 'subsets sleep for $n_s$ slots', and 'without subsets' approaches, according to the number of rounds, is shown in Fig. 15. The "without subsets" approach in Fig. 15 also refers to the SENDORA network and the CRN with the LEACH-C protocol. Similar to the SENDORA network and the CRN with the LEACH-C protocol, the 'without subsets' approach utilizes all sensor nodes in a cluster to perform spectrum sensing. In all the approaches, identical cluster is associated with each CR. As seen in Fig. 15, mean squared error by the proposed algorithm ("subset with max energy") is a little increased from the ones by the other works ("without subsets"), whereas the energy consumption is significantly increased with the other works, as observed in Fig.11, and the end-to-end delay is escalated with the other works, depending on the architecture of the sensor network, as shown in Fig.10. It is evident from the figure that the 'subsets sleep for $n_s$ slots' approach produces more error compared with its counterparts. That is because $n_s$ is estimated from historical behavior of the PU, so it cannot predict future behavior of the PU accurately all the time. However, the error penalty (loss due to error) is compensated by the reward of reduced energy consumption, which is demonstrated in Fig. 12. On the other hand, other approaches incur comparatively few errors because there is at least one active subset in each cluster.

Figure 16 shows the global detection probability. Since the minimum global detection probability is 0.8, as given in Table 1, it is obligatory for all the approaches to satisfy this minimum threshold while reducing energy consumption and MSE. The figure shows that all the approaches satisfy the minimum global detection probability. It is seen that the detection probabilities of the 'subset with max energy' approach and 'subsets sleep for $n_s$ slots' approach are greater than the minimum global detection probability with the reduced energy consumption.

## VII. CONCLUSION

In this paper, an ad hoc CRN with an energy-efficient process, namely the CUSF process, is proposed. Via the CUSF process, clustering and further subset formation of the sensor nodes are performed. Multiple subsets are created in a cluster and only one subset is active in sensing to reduce energy consumption. For further reduction of energy consumption, the actual sensor nodes for spectrum sensing are selected in the given active subset according to a separately proposed algorithm. In addition, all the subsets, including the one active subset, switch to sleep mode for the duration of PU activity to achieve another reduction in energy consumption. A novel subset scheduling algorithm to achieve this goal is developed on the basis of PU statistics. As a result, the CRN with the proposed architecture consumes significantly less energy and incurs lower end-to-end delay in comparison with the SENDORA network and the CRN with the LEACH-C protocol.